\documentclass[dvips]{article}
\usepackage{icrctc07}

\title{The Baikal Neutrino Telescope: Status and Plans}
\shorttitle{Baikal Neutrino Telescope: Status and Plans}

\authors{K. Antipin$^{1}$, V. Aynutdinov$^{1}$, V. Balkanov$^{1}$, 
I. Belolaptikov$^{4}$, N. Budnev$^{2}$, I. Danilchenko$^{1}$, G. Domogatsky$^{1}$,
A. Doroshenko$^{1}$, A. Dyachok$^{2}$, Zh. Dzhilkibaev$^{1}$, S. Fialkovsky$^{6}$,
O. Gaponenko$^{1}$, K. Golubkov$^{4}$, O. Gress$^{2}$, T. Gress$^{2}$, O. Grishin$^{2}$, A. Klabukov$^{1}$,
A. Klimov$^{8}$, A. Kochanov$^{2}$, K. Konischev$^{4}$, A. Koshechkin$^{1}$, V. Kulepov$^{6}$, L. Kuzmichev$^{3}$,
E. Middell$^{5}$, S. Mikheyev$^{1}$, M. Milenin$^{6}$, R. Mirgazov$^{2}$, E. Osipova$^{3}$, Yu. Pavlova$^{1}$,
G. Pan'kov$^{2}$,
L. Pan'kov$^{2}$, A. Panfilov$^{1}$, D. Petukhov$^{1}$, E. Pliskovsky$^{4}$, P. Pokhil$^{1}$, V. Poleshuk$^{1}$,
E. Popova$^{3}$, V. Prosin$^{3}$, M. Rosanov$^{7}$, V. Rubtzov$^{2}$, B. Shaibonov$^{4}$, A. Sheifler$^{1}$,
A. Shirokov$^{3}$, Ch. Spiering$^{5}$, B. Tarashansky$^{2}$, R. Wischnewski$^{5}$, I. Yashin$^{3}$,
V. Zhukov$^{1}$
}
\shortauthors{Author and et al.}
\afiliations{$^1$Institute for Nuclear Research, Moscow, Russia\\ 
$^2$Irkutsk State University, Irkutsk, Russia\\
$^3$Skobeltsyn Institute of Nuclear Physics  MSU, Moscow, Russia\\
$^4$Joint Institute for Nuclear Research, Dubna, Russia\\
$^5$DESY, Zeuthen, Germany\\ 
$^6$Nizhni Novgorod State Technical University, Nizhni Novgorod, Russia\\ 
$^7$St Petersburg State Marine University, St Petersburg, Russia\\ 
$^8$Kurchatov Institute, Moscow, Russia\\ 
}
\email{ralf.wischnewski@desy.de}

\abstract{The high energy neutrino telescope NT200+ is 
currently in operation in Lake Baikal. We review the status of 
the Baikal 
Neutrino Telescope,
and describe recent progress on key components of the 
next generation kilometer-cube (km3) Lake Baikal 
detector,
like 
investigation 
of new large area phototubes, 
integrated
into the 
telescope.
}

\begin{document}
\maketitle
\section{Introduction}

The Baikal Neutrino Telescope  is operated in Lake
Baikal, Siberia,  at a depth of {1.1~km}.
Deep Baikal water is characterized by an absorption length of 
$L_{abs}(480 $nm$) =20\div 24$ m,
a (geometric) scattering length of $L_s =30\div 70$ m and 
a strongly anisotropic scattering function
with a mean cosine of scattering angle $0.85\div 0.9$ \cite{APP1},
and by 
a level of bioluminescence and other natural 
backgrounds that are well below seawater sites.

The first stage telescope, NT200, started full operation in spring 1998
and contained 192 Optical Modules (OMs).
The favorable water properties, and a relatively simple and 
reliable design led to the physics success of this 
comparably small telescope.
Low light scattering allows for a sensitive volume of a few Mtons at 
PeV shower energy scale,
well
beyond the geometric detector limits.
For a review of the high sensitivity limits on UHE astrophysical 
neutrino's as well as best so far obtained limits on relativistic 
magnetic monopoles and other results,  
see \cite{ICRC07_B}.
The upgrade to NT200+ was a logical consequence of the 
large external sensitive volume, now to be fenced by 
sparsely
instrumented 
external strings of OMs.

In this paper, we review the current status of the 
Baikal Neutrino Telescope 
as of
2007,
and the 
activities towards the km3-scale detector. 
Results on a prototype device for acoustic neutrino detection,
obtained with a stationary setup in 2006/2007, are reported 
elsewhere in these proceedings \cite{ICRC07_C}.

\section{The NT200+ Telescope}


\begin{figure}[t]
\begin{center}
\includegraphics [width=0.4\textwidth]{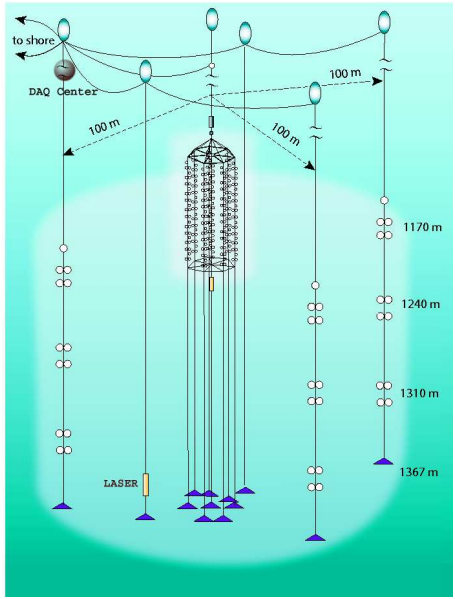}
\caption{
Baikal Telescope NT200+ :
old NT200 surrounded
by three external long strings;
indicated: external laser and underwater DAQ center.
\label{FIG_NT200}}
\end{center}
\vspace*{-7mm}
\end{figure}

The 
telescope NT200+ \cite{CAT1}
was
commissioned in April, 2005,
and is made of 
a central part (the old telescope NT200)
and three additional 
external strings, 
see Fig.\ref{FIG_NT200}.
Underwater electrical cables connect the
detector with the shore station.

The 
first stage telescope configuration  NT200
\cite{APP1} 
is made of 
an umbrella-like frame, 
carrying 8 strings,
each with 24 pairwise arranged OMs
(see central part of Fig.\,\ref{FIG_NT200}).
Each optical module
contains a 37-cm diameter photomultiplier (PM) 
{QUASAR-370},
developed specially for this project
\cite{OM2}.
The two PMs of a pair are switched in local coincidence 
(a {\it channel})
in order
to suppress background from bioluminescence and PM noise.

The external strings 
of NT200+
are 200~m long (140~m instrumented)
and are placed at 
100~m
distance from the  center of NT200. 
Each string contains 12\,OMs, also pairwise grouped
like in NT200.
The upper channels are at approximately the same depth as the bottom OMs 
of NT200,
adjacent channel distances are 20, 50, 20, 30 and 20\,m from top to bottom
(for absolute depths of upper, 3rd and lower channels see Fig.\ref{FIG_NT200}).
All channels are
downlooking,
except the lower two on each string (uplooking).

Changes made to NT200+ during the 2007 
expedition are
described below; 
in addition a half-string of NT200
was lowered by 85\,m, to
improve background rejection and lower the shower threshold.




\begin{figure}[t]
\vspace*{-0.3cm}
\begin{center}
\includegraphics [width=.45\textwidth]{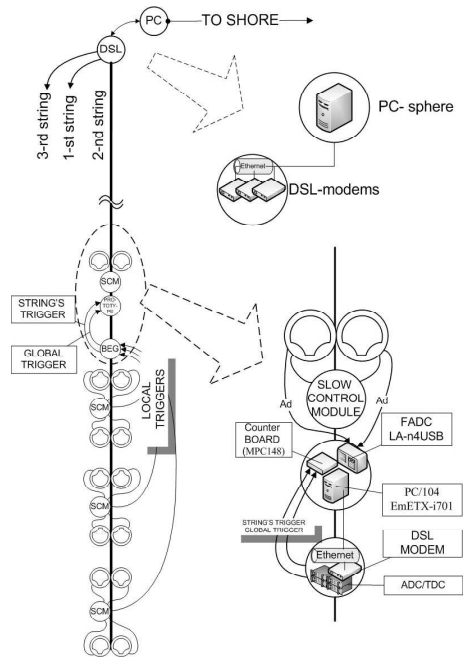}
\vspace*{-0.7cm}
\caption{
The external 
string 
readout/control 
system,
with the 
FADC prototype 
for two 13" PMs.
\label{FIG_FADC}}
\end{center}
\vspace*{-8mm}
\end{figure}


\begin{figure}[t]
\begin{center}
\includegraphics [width=.40\textwidth]{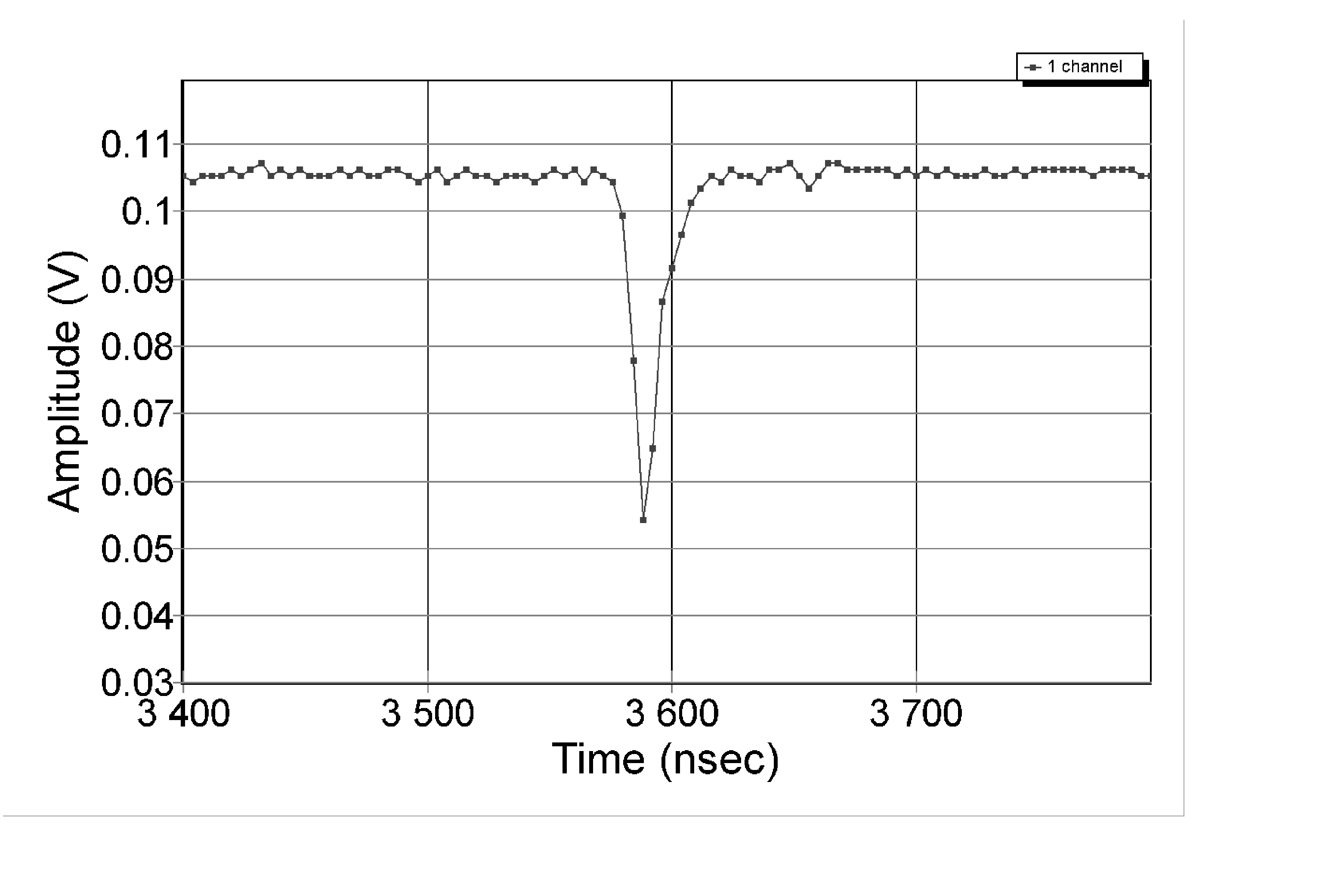}
\vspace*{-0.4cm}
\includegraphics [width=.40\textwidth]{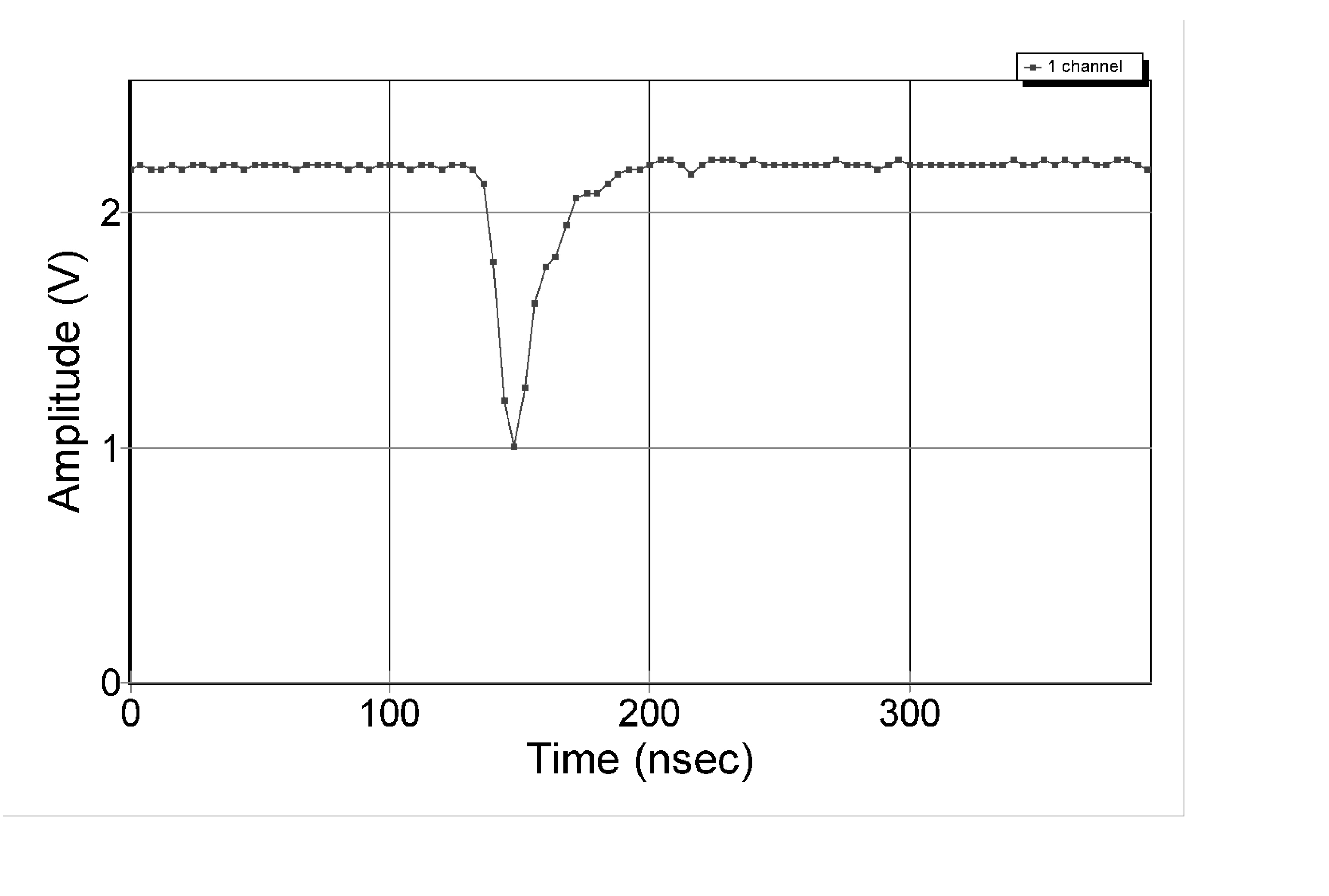}
\vspace*{-0.4cm}
\includegraphics [width=.40\textwidth]{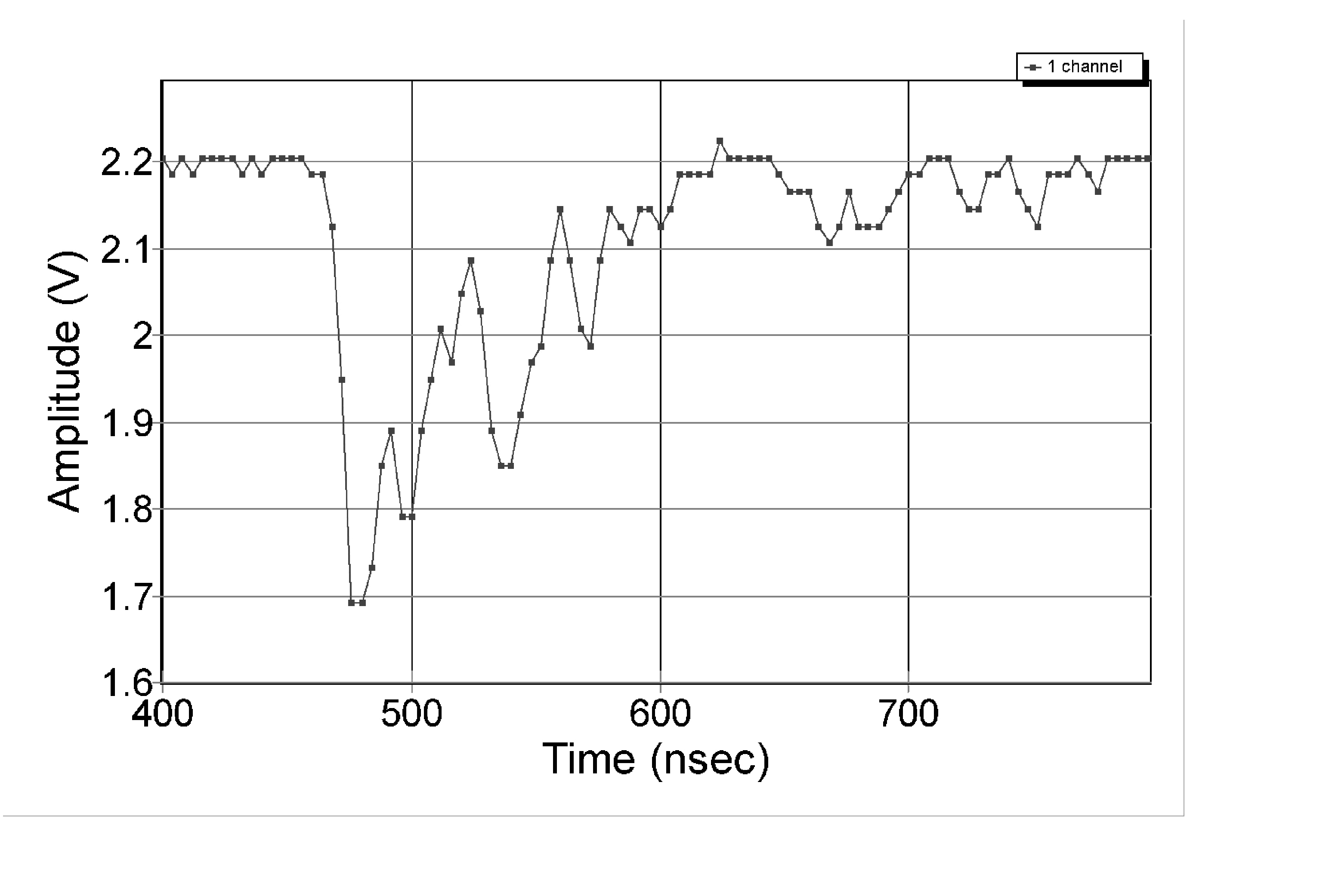}
\vspace*{+0.2cm}
\caption{
Examples of 
13''\,PM pulses
from FADC-prototype, see fig.\ref{FIG_FADC}.
Upper: noise pulse,
middle: muon event, 
lower: 
laser backward illumination.
\label{FIG_SOME_PULSE}}
\end{center}
\vspace*{-5mm}
\end{figure}


\begin{figure}[t]
\begin{center}
\includegraphics [width=.60\textwidth]{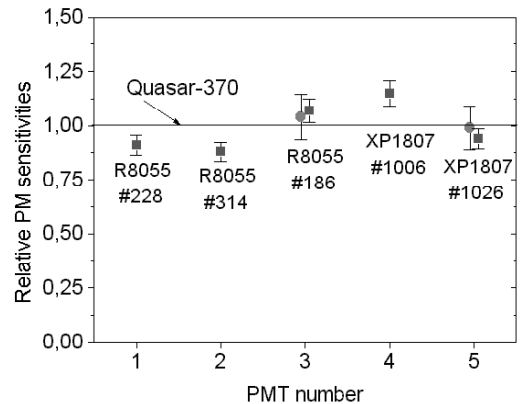}
\vspace*{-2.0cm}
\caption{
Ratio of effective sensitivity of 
large 
area 
PMs 
R8055/13" and XP1807/12" 
to QUASAR-370/14.6".
Laboratory (squares), in-situ (dots).
\label{FIG_PMTS}}
\end{center}
\vspace*{-.50cm}
\end{figure}



\section{Towards a km3 detector in Lake Baikal}

The construction of 
NT200+ is a first step towards a km3-scale Baikal 
neutrino telescope.
Such a detector could be made of 
building blocks similar to NT200+, but
with NT200 replaced by a single string,
still 
allowing separation of high energy neutrino cascades
from background.
%
It will contain 
a total of 1300-1700 OMs,
arranged at 
90-100 strings with 12-16 OMs each, and a length of 300-350\,m.
Interstring distance will be $\approx$100\,m. 
The effective volume for cascades with energy above 100\,TeV  
is 0.5-0.8\,km$^3$.


The existing NT200+ 
%
allows to 
verify 
all
key elements 
and 
design principles 
of the km3 (Gigaton-Volume) Baikal telescope. 
%
Next milestone of the ongoing  
km3-telescope research and development work (R\&D) 
will be spring 2008: 
Installation of a ``new technology'' prototype string,
as part of NT200+.
This string will consist of 16 optical modules and an FADC based
measuring system.
%
Three issues, discussed in the remainder of this paper,
are 
investigated 
in 2007, and 
will permit installation of this prototype string:
(1) increase of underwater (uw) data transmission bandwidth,
(2) in-situ study of FADC PM-pulses,
(3) preliminary selection of optimal PM.

\subsection{Modernization of data acquisition system}

The basic goal of the NT200+ DAQ modernization is 
a substantial increase of the uw-data rate -
to allow for transmission of significant FADC data rates,
and also for more complex trigger concepts 
(e.g. lower thresholds).
In a first step, in 2005 a high speed data/control tcp/ip connection between the shore
station and the central uw-PCs (data center) had been established 
(full multiplexing over a single pair of wires, with a hot spare) \cite{ICRC05_DAQ}, 
based on DSL-modems (FlexDSL).   
In 2007, 
the communication on the remaining segment uw-PC - string controller was upgraded 
using the same approach, 
see fig.\ref{FIG_FADC}.
The basic elements are new string-controllers (handling 
TDC/ADC-readout)  
with an ethernet-interface, connected by a 
DSL-modem to 
the
central
uw-DSL unit (3 DSL modems, max. 2\,Mbps each), connected by ethernet to the uw-PCs. 
The significant increase in uw-data rate (string to uw-PC)
provided the possibility to 
operate the new prototype FADC system.
%
%
%

\subsection{Prototype of a FADC based system}

A prototype FADC readout system was installed during the Baikal expedition 2007.
It should yield 
input for the design of the 2008 km3-prototype string (FADC), 
such as: 
optimal sampling time window, dynamic range, 
achievable 
pulse parameter precisions, 
algorithms for online data handling, estimation of true bandwidth needs.
These data will also be useful
to decide about the basic DAQ / Triggering approach for the km3-detector:
at this stage, both a complex FADC based, as well as a classical TDC/ADC
approach seem feasible.

The FADC prototype 
is located at the 
upper part of the 2nd outer string, see Fig.\ref{FIG_FADC}. 
It includes two optical modules with up-looking PM R8055, 
a slow control module and a FADC sphere. 
The FADC sphere consists of two 250 MSPS FADCs, with 
USB connection to an embedded PC104 computer emETX-i701, 
and a counter board MPC148. 
The standard string trigger (2-fold channel coincidence) 
is used as FADC trigger.
Data are transfered via local ethernet and the DSL-link of the 2nd string. 
%
Data analysis from FADC the prototype is in progress.
Fig.\ref{FIG_SOME_PULSE} 
shows examples of FADC pulses 
for different classes of events.
The upper panel gives a 1 photoelectron (p.e.) noise hit, for scale.
A muon trigger (multi-p.e.) is given in the middle panel;
the lower panel shows 
an interesting event, due to  
backward illumination by an intense calibration laser,
located $\sim$140\,m away.
The PM orientation opposite to the calibration laser explains 
the significant signal duration ($>$\,100\,ns),
illustrating the light
scattering influence on 
particle detection for 
large distances.

\vspace*{-2mm}
\subsection{PM selection for the km3 prototype string}

Selection of the optimal PM type for the 
km3 telescope
is a key question of detector design. 
Assuming similar values for time resolution and linearity range,
the basic criteria of PM selection is its 
effective sensitivity to Cherenkov light,
determined as the fraction of registered photons per 
photon flux unit.
It is 
determined  by 
photocathode area, 
quantum efficiency, 
and photoelectron collection efficiency.
We compared effective sensitivities 
of Hamamatsu 
R8055 (13" photocathode diameter) and XP1807
(12")
with
Quasar-370 (14.6") \cite{OM2},
which was successfully operated in NT200
over more than 15 years. 
In laboratory we used blue LEDs (470 nm),
located at 150\,cm distance from the PM.
Underwater measurements are done for 
2 R8055 and 2 XP1807, installed permanently as two NT200-channels,
which are illuminated by 
the external laser calibration source \cite{ICRC05_DAQ}, located 
$160-180$\,m away (see Fig.\ref{FIG_NT200}).
Preliminary results of these 
effective PM sensitivity measurements 
are given in Fig.\ref{FIG_PMTS},
and show relatively small deviations.
Smaller size (R8055, XP1807) 
tends 
to be compensated by larger photocathode sensitivties. 
In addition, we 
emphasize the advantage of a spherical shape (as QUASAR-370);
we are investigating the angular integrated sensitivity losses due to 
various deviations from that optimum.

\vspace*{-3mm}
\section{Summary}

The Baikal Neutrino Telescope is taking data currently
in it's NT200+ configuration - 
an upgrade
of the original NT200 telescope
for improved high energy shower sensitivity.

%

For a
km3-detector in Lake Baikal, 
R\&D-activities have been started.
The NT200+ detector is, beyond its better physics sensitivity, 
used as an ideal testbed for 
critical new components.
Modernization of the NT200+  DAQ 
allowed to install a prototype 
FADC PM readout.
Six large area hemispherical PMs have been 
integrated 
into 
NT200+
(2 Photonis XP1807/12" and 4 Hamamatsu R8055/13"), 
to facilitate
an optimal PM choice.
A 
prototype 
new technology string will be installed in spring 2008;
and a km3-detector Technical Design Report 
is planned for fall 2008.



{\it This work was supported by the Russian Ministry of Education and Science, the
  German Ministry of Education and Research and the Russian Fund of Basic Research
  (grants 05-02-17476, 05-02-16593, 07-02-10013, 07-02-00791), by the Grant of
  the President of Russia NSh-4580.2006.2 and by NATO-Grant
  NIG-9811707 (2005).}

\end{document}